# LLMbench: A Comparative Close Reading Workbench for Large Language Models

David M. Berry

d.mberry@sussex.ac.uk

University of Sussex

## Abstract

LLMbench is a browser-based workbench for the comparative close reading of large language model (LLM) outputs. Where existing tools for LLM comparison, such as Google PAIR's LLM Comparator are engineered for quantitative evaluation and user-rating metrics, LLMbench is oriented towards the hermeneutic practices of the digital humanities. Two model responses to the same prompt are side by side in annotatable panels with four analytical overlays (Probabilities for token-level log-probability inspection, Differences for word-level diff across the two panels, Tone for Hyland-style metadiscourse analysis, and Structure for sentence-level parsing with discourse connective highlighting), alongside five analytical modes, Stochastic Variation, Temperature Gradient, Prompt Sensitivity, Token Probabilities, and Cross-Model Divergence, that make the probabilistic structure of generated text legible at the token level. The tool treats the generated text as a research object in its own right from a probability distribution, a text that could have been otherwise, and provides visualisations including continuous heatmaps, entropy sparklines, pixel maps, and three-dimensional probability terrains, that show the counterfactual history from which each word emerged. This paper describes the tool's architecture, its six modes, and its design rationale, and argues that log-probability data, currently underused in humanistic and social-scientific readings of AI, is an important resource for a critical studies of generative AI models.

*The meaning of a text lies not behind the text but in front of it. The meaning is not something hidden but something disclosed.*

Paul Ricoeur, 1981.

The proliferation of large language models (LLMs) has generated an equally prolific effort to measure them. Kahng et al. (2024) name part of the problem directly in their discussion of Google PAIR's LLM Comparator. Side-by-side evaluation of models, they argue, is a key practice, and existing tools for it tend to be quantitative or rely on presenting user-rating metrics. The system they present is a useful piece of engineering, but it is designed for model developers (especially those making products), not for the hermeneutic work of close reading what a model has generated. What is missing is a workbench for comparative close reading, an environment in which the outputs of two models can be subjected to the kinds of attention the humanities bring to primary sources, annotation, structural differences, rhetorical analysis, or, especially in relation to LLMs, inspection of the probabilistic structure from which the text emerged. This paper introduces a new comparative tool for undertaking close reading of LLM outputs side-by-side called LLMbench, which offers a number of innovative methods for researchers trying to understand the outputs of these systems.[1]

The digital humanities (DH) have a long tradition of tool building for textual analysis, Voyant Tools (Sinclair and Rockwell 2016) probably being the most familiar, alongside TAPoR, AntConc, MALLET, CATMA, Recogito, and a lineage of corpus-reading environments designed for scholarly rather than evaluative work. These tools count, visualise, index, and annotate, and they support both close and distant reading. They were built for a fixed object, a human-authored text or corpus whose meaning is explored through structural and statistical description. What they were not built to do is treat the text itself as a probabilistic object (i.e. a draw from a probability distribution that could just as easily have given different tokens at the same position). LLMbench inherits the DH tradition of browser-based, annotation-rich tool building, but the hermeneutic problem it takes on is different, how to read a text that could have been otherwise, and how to read it alongside another text that could have been otherwise in different ways.

LLMbench, part of what I call Vector Lab tools, that are built to make those differences readable. It has six modes. The Compare mode is the primary close-reading area, where two model responses to the same prompt sit side by side in annotatable panels with four analytical overlays. In addition to this are five Analyse modes that can be used to run empirical probes, (1) Stochastic Variation, (2) Temperature Gradient, (3) Prompt Sensitivity, (4) Token Probabilities, and (5) Cross-Model Divergence. The figures that follow use a prompt about Italo Calvino's *Cybernetics and Ghosts*, comparing the models Gemini 2.0 Flash against GPT-4o.[2]

---

[1] LLMbench is built with Next.js 16 and React 19, deployed on Vercel, with the Vercel AI SDK providing a unified interface across the supported providers (Anthropic, OpenAI, Google, Hugging Face, and OpenRouter). The prose display panels use CodeMirror 6, a code-editor substrate that provides the gutter, line-based rendering, and annotation widget infrastructure. Styling is Tailwind CSS with a custom editorial palette. API keys and comparison history persist to browser localStorage; no user data is sent to any server beyond the model providers themselves. The full source is MIT licensed at https://github.com/dmberry/LLMbench.

## Related Work

The comparison of LLM outputs is currently dominated by what might be called the engineering evaluation tradition. Kahng et al.'s (2024) LLM Comparator, developed at Google PAIR, is representative. It assumes a fixed rubric, a set of prompts, and a pair of models to be ranked according to win-rates, safety violations, or user-rating metrics. The interface shows this through bar charts, leaderboards, and rationale-summarisation features that abstract from the particulars of any given output. LMSYS Chatbot Arena works at scale through crowdsourced pairwise preferences, generating Elo ratings that are then tabulated in public leaderboards (Chiang et al. 2024). HELM and BIG-bench extend the logic across hundreds of tasks and dozens of metrics, producing exhaustive tabulations of model performance (Liang et al. 2023; Srivastava et al. 2023). These are useful instruments. They are not, and do not claim to be, environments for close reading. Their analytical object is the model, not the text the model produces.

The digital humanities tradition of tool-building occupies a different problem space. Voyant Tools (Sinclair and Rockwell 2016) is probably the most familiar, offering word frequencies, collocation networks, trend lines, and reader views for exploring a fixed corpus. TAPoR served as a directory and portal for text analysis tools for many years. AntConc (Anthony 2022) provides concordancing and keyword extraction for corpus linguistics. MALLET (McCallum 2002) remains a workhorse for topic modelling. CATMA (Gius et al. 2021) and Recogito (Simon et al. 2017) support scholarly annotation, the former for textual markup and the latter for geographical and entity-level tagging. The lineage is long, and includes earlier environments like WordHoard, TuStep, and the corpus tools of the Text Encoding Initiative (TEI). What unites these is an assumption about their object. The text is given and it is a relatively stable object. What varies is the reader's attention, the statistical description, the annotation scheme applied to it. Variation within the text through probability distributions is not an analytical category that is used often, or where it is, as in variorum editions, it is a matter of the comparison of historical materials rather than of probability distributions over tokens at time of generation.

The gap between these two traditions is where LLMbench is positioned. Where the engineering tools treat the text as evidence for claims about the model, and the DH tools treat the text as a quasi-fixed object for critical work, LLMbench treats the text as a probabilistic object that could just as easily have given different tokens at the same position. This is a substantively different hermeneutic situation. It demands tools that make the distribution itself legible, not only the realised text. Log-probabilities (i.e. logprobs), already routinely returned by commercial model APIs, provide the raw material for this new tool. The challenge has been how to render that material readable alongside the text, how to stage comparison across distributions produced by different models, and how to do so without losing a sense of the hermeneutic approaches of the humanities.

---

[2] The twelve figures above are screenshots from the LLMbench tool. The prompt used throughout is "Tell me about Calvino's *Cybernetics and Ghosts* and its relevance for AI today," with Gemini 2.0 Flash in Panel A and GPT-4o in Panel B. The same prompt and pairing run across all figures so that patterns across modes can be read as belonging to a single comparison rather than as separate examples.

Beyond these two traditions, some adjacent work is worth naming. Critical Code Studies (Marino 2020; Berry and Marino 2024) extends close reading from literary text to source code, arguing that code is a cultural text whose meaning emerges through the interplay of technical function, authorial choice, and social context. CCS offers an annotation methodology (Observation, Question, Metaphor, Pattern, Context, Critique) that LLMbench inherits and adapts for model outputs. CCS reads code that has been written, with its layers of intention and convention. Reading LLM output is reading a text that was generated, where the question of intention is different and where the probabilistic structure of generation is itself a hermeneutic object. Alongside this, the critical reception of generative AI within media and cultural studies (Pasquinelli 2023; Berry 2025) has mainly proceeded through theoretical analysis rather than tool-building. LLMbench is an attempt to provide an instrument that might support this kind of critical reading empirically, at the level of the generated text.

The variorum principle from *10 PRINT* (Montfort et al. 2013) offers a useful resource for thinking about this. Reading a one-line BASIC program through the many variants it could produce, the authors show how a simple probabilistic structure generates an analytically productive surface. LLMbench extends this move to the very different object of large language model output. Different runs of the same prompt, different models responding to the same prompt, different temperature settings, are variants whose comparison reveals what is deterministic and what is contingent in the generated text. The tool makes those variants readable side by side.

## Design Rationale

Tools can easily shape the readings they enable. A workbench built for evaluating models encourages evaluative reading. A workbench built for close reading of a fixed corpus encourages that kind of attention. LLMbench takes a particular position on what reading an LLM output involves, and the tool's architecture follows from that position.

A generated text, at the moment it is sampled, is the realisation of a probability distribution over tokens. For each position in the sequence, the model has a full ranked list of candidate tokens weighted by probability, and the sampler selects one. What appears to the reader as a finished sentence is the trace of a sequence of such selections. It is the mechanism of generation, and usefully, commercial model APIs increasingly return log-probability data that exposes it. Log-probabilities are numeric, per-token, and carry with them the top-k alternatives the model considered at each position. They make the probabilistic structure of the text directly inspectable.

Reading this kind of text requires a different hermeneutic orientation from reading human-authored prose. Close reading of, say, a poem works on the text as a finished object, where every word is the one chosen and the question is what that choice means. Close reading of LLM output can proceed this way too, treating the generated text as if it were finished and asking what it means. Doing so, however, discards what is most distinctive about the object. The generated text is not only what appears on the page. It is one taken from a distribution of near-equivalent alternatives, many of which were weighted nearly as highly. To read the text as if the specific word at position 26 were the inevitable word at position 26 is to read past the probabilistic structure that defines the text as an LLM output.

What I am calling *comparative AI hermeneutics* names a reading practice that holds both dimensions in view at once. It reads the realised text as text, attending to its rhetorical moves, its structural choices, its argumentative shape. It also reads the distribution the realised text was drawn from, attending to where the model was confident, where it was uncertain, where it forked, where another plausible word would have taken the passage in a different direction. The two readings are not in competition. They surface different features of the same object, and the LLMbench enables the reader to move between them.

Comparison across models extends the practice further. When two models respond to the same prompt, the comparative reading is not only between two realised texts, it is between two distributions. Where one model is confident, the other may be uncertain. Where one model forks along a particular line, the other may commit firmly. These differences in the shape of uncertainty are as analytically interesting as the differences in the surface text, and they are not visible at the surface. The tool's four analytical overlays (i.e. Probabilities, Differences, Tone, Structure) attempt to make both layers readable side by side.

The annotation methodology borrows from Critical Code Studies (Marino 2020; Berry and Marino 2024), which has spent over a decade developing an approach to reading source code as cultural text. The six annotation categories used in LLMbench are drawn from the CCS methodology as elaborated in the ELIZA reading project and CCS work group discussions. These categories were not developed for LLM output. CCS developed them as basic moves of scholarly reading applied to texts that are not literary in the usual sense, and those moves transfer usefully to the hermeneutic challenges of reading generated prose.

A final design choice concerns the reading user interface. The panels in LLMbench are implemented in CodeMirror, a code-editor framework rather than a word-processor framework. This is a decision made by the sibling CCS Workbench and inherited here. Displaying LLM prose in a code-editor substrate produces a mild defamiliarisation, the text looks slightly less like natural language and slightly more like an object to be analysed. The gutter-based annotation system that CodeMirror supports is also a better fit for scholarly annotation than the inline-comment conventions of word processors. The UI is itself part of the reading practice the tool is built to enable.

## Token Probabilities in Compare Mode

The Probs overlay is the analytical heart of LLMbench.[3] Activating "Probs" in the Compare toolbar re-sends the current prompt to the model API and requests what is called "logprob" data alongside the response. This takes a few seconds. Once loaded, a continuous heatmap overlays both panels. The idea is simple. Tokens the model chose with high confidence (above roughly 70% probability) are given no highlight. Below that threshold the background colours progressively, from pale yellow through orange to deep red for positions where the model was "uncertain". The full probability gradient is visible across both panels at once, so confidence patterns in each response can be compared.

The navigation strip below the toolbar provides more analytical controls. Three buttons summarise the data at a glance. "Uncertain" jumps to the positions with the highest entropy, where the model was most genuinely uncertain across alternatives (i.e. where no single token was clearly favoured). "Forks" jumps to positions where the chosen token had less than 70% probability, a lower bar that surfaces a broader population of possibilities. "Diverge", available when both panels have logprob data, jumps to positions where Panel A and Panel B chose different tokens at the same sequence position.

In the Calvino comparison shown in Figure 1, those numbers show 399 uncertain positions, 174 forks, and 281 diverge points across the two responses.

---

[3] Logprob data is available from Google Gemini (version 2.0 and later), OpenAI models directly, and OpenAI models via OpenRouter. GPT-4o and GPT-4o-mini return reliable logprob data through OpenRouter. Other models hosted on OpenRouter do not currently expose choice.logprobs through the routing layer. Selected Hugging Face models via Fireworks and Together inference backends also support logprobs. A logprobs-compatible only checkbox in Provider Settings greys out providers and models that do not expose token probabilities.

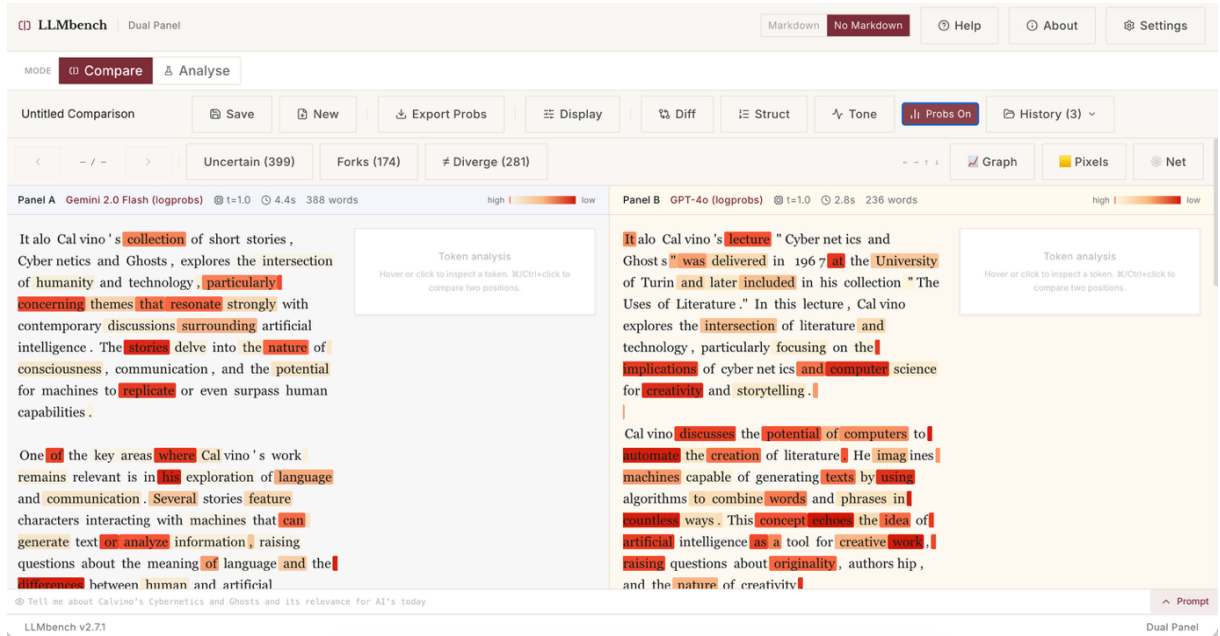

Figure 1: Compare mode with Probs overlay active across both panels. The continuous token heatmap is visible on each panel, ranging from uncoloured (confident) through yellow and orange to deep red (uncertain). The navigation strip is visible below the toolbar, showing Uncertain (399), Forks (174), and Diverge (281) counts. Both panels display Gemini 2.0 Flash and GPT-4o responses to the Calvino prompt, with the Tone analysis side panel on each side.

The user can click on a token and a probability distribution panel opens alongside the text, showing the full top-k alternatives the model was considering at that position. This is where the reading becomes interesting. The inspector shows both the chosen token and the probability mass the model assigned to everything it did not choose. At position 26 in the Gemini response, for instance, the model was working with an entropy of 2.315 bits and chose its token with only 11.76% probability, meaning several other tokens were live and nearly as probable. The GPT-4o response at the same prompt position shows a very different distribution, entropy of 1.567 bits, chosen probability 49.27%. The two models are not equally uncertain at the same moment. They encounter different distributions of difficulty as they traverse the same conceptual territory, and the inspector makes that visible at the token level. The user can command or control-click on a second token to pin two distributions side by side, enabling direct comparison of two moments of uncertainty within the same response, or of the same position across both panels.

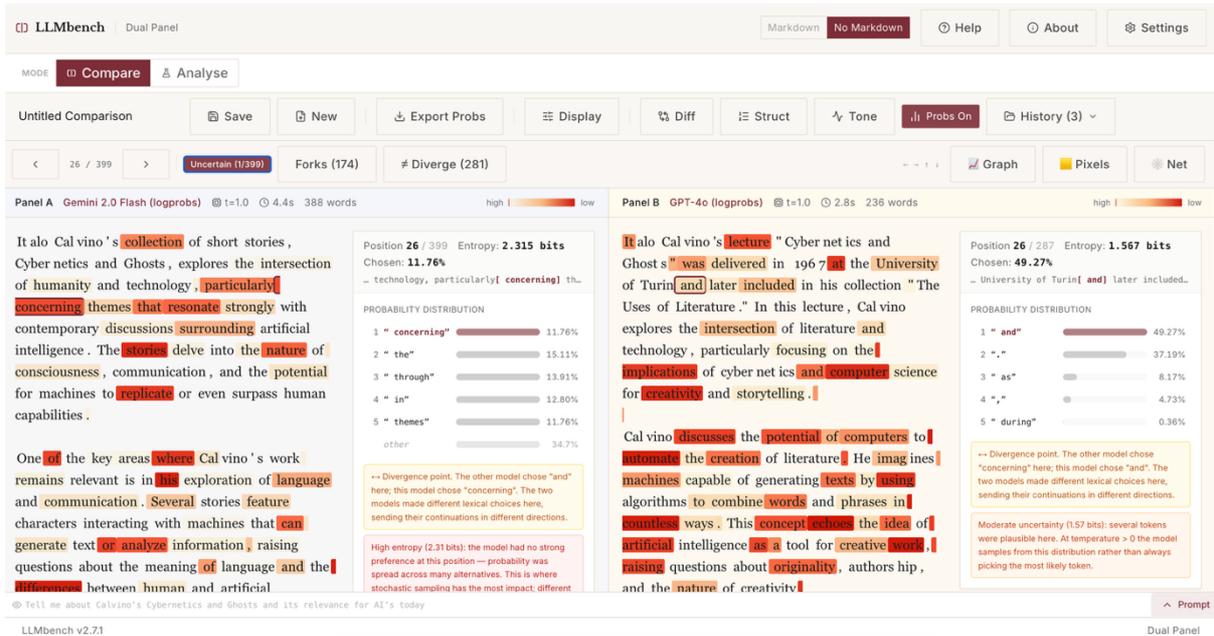

Figure 2: The probability inspector panel pinned in both Panel A and Panel B. Panel A shows Position 26/399, Entropy 2.315 bits, Chosen 11.76%, with probability distribution bars for the top alternatives. Panel B shows Position 26/287, Entropy 1.567 bits, Chosen 49.27%, with its own distribution. The divergence annotation is visible below each inspector, noting that the two models chose different tokens here.

Three optional visualisation bands extend the Probs analysis beyond the inline heatmap. Each represents the same underlying probability data through different spatial visualisations, and each makes different patterns more visible.

The "Graph" band renders an entropy curve, an SVG sparkline of per-token entropy plotted across the entire sequence, with Panel A and Panel B overlaid in different colours. The horizontal axis is token position, the vertical axis entropy in bits. Reading the curve reveals the rhythmic structure of each model's uncertainty, where it was consistently committed across a passage, where it suddenly became spread, and crucially, where the two models diverge in their entropy profiles despite having received the same prompt. The user can click any point on the curve to jump the cursor in both panels to that token position.

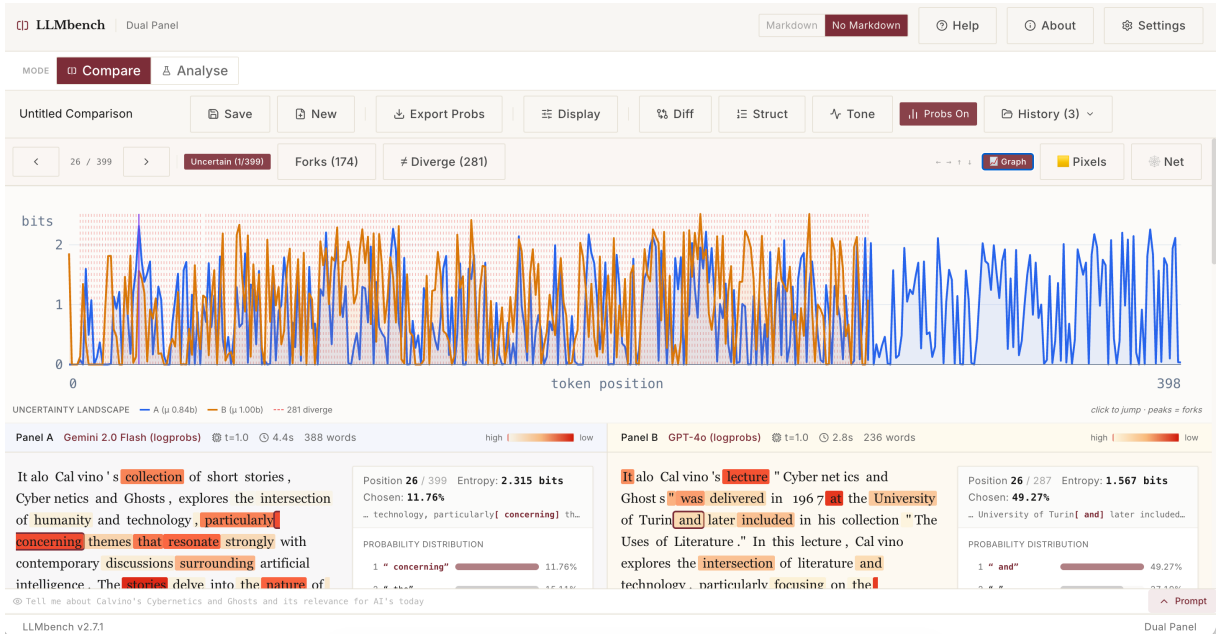

Figure 3: The Graph band active, showing the entropy curve at the top of the interface. Blue and orange sparklines (Panel A and B respectively) trace per-token entropy across all token positions, with the horizontal axis showing token position up to approximately 398 and the vertical axis measuring bits (0 to 2). Below the curve, both panel heatmaps remain visible with the probability inspector open. The Uncertain and Diverge navigation chips are active in the strip.

The "Pixels" band offers something different, a bird's-eye summary of both responses at once. Each token becomes a coloured cell in a dense grid, with the same heat palette as the inline heatmap. Where the heatmap embedded in the prose requires reading to navigate, the pixel map collapses the entire response into a single glance-level view. Clusters of red cells indicate passages of sustained uncertainty, whereas pale or uncoloured expanses indicate confident stretches. Because both panels use the same cell size, the spatial extent of each response is immediately comparable, and the distribution of uncertainty visible across each grid can be read against the other.

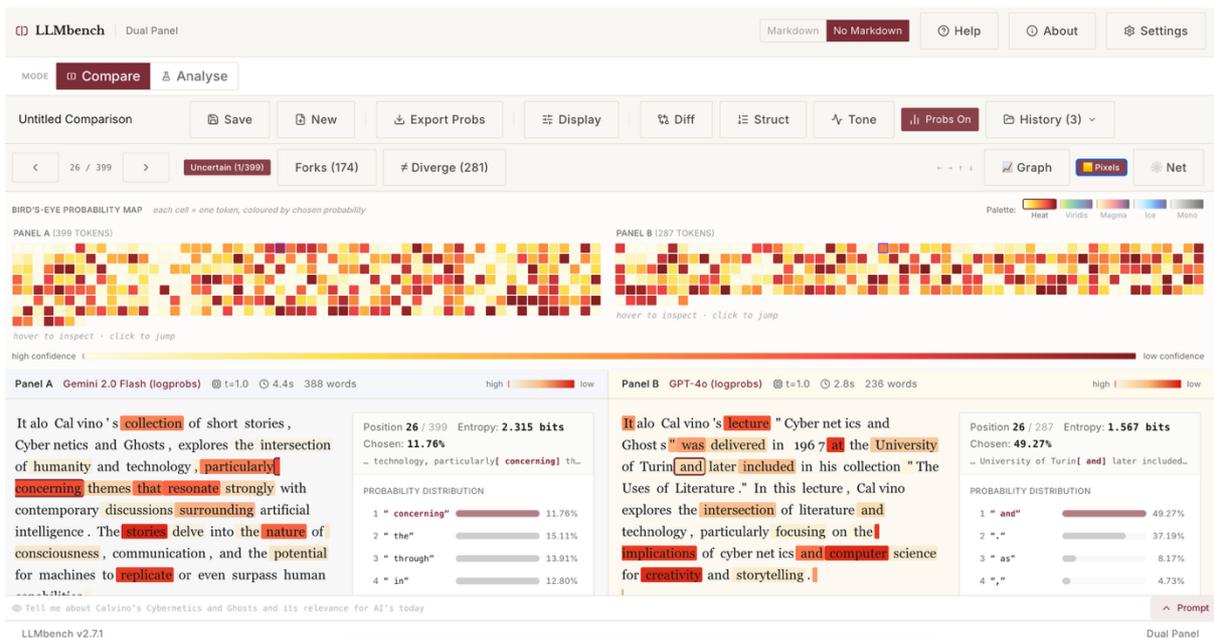

Figure 4: The Pixels band active, showing the token pixel map for both panels above the regular text view. Each cell represents one token, coloured by probability using the Heat palette. Panel A (Gemini 2.0 Flash, 398 tokens) and Panel B (GPT-4o, 267 tokens) are displayed side by side, with their different response lengths visible in the different widths of the grids. The colour distribution varies noticeably between the two panels, particularly in the density of red cells.

The "Net" band renders the probability data as a three-dimensional terrain. Each token position becomes a vertex on a mesh surface and the vertical displacement of each vertex corresponds to its entropy. High-entropy positions become peaks and confident stretches become flat plains. The mesh is translucent with a wireframe net overlay and can be rotated freely by dragging. The top five highest-entropy peaks carry floating labels showing the token text and its entropy value. The user can click any point on the surface to jump the cursor to that position in both panels.

The three bands, curve, pixels, and net, give different ways to visualise the probability data. The curve emphasises temporal dynamics, how uncertainty evolves across the sequence as a linear narrative. The pixel map offers a glance-level spatial summary. The net turns uncertainty into navigable terrain that can be inspected from any angle. Each foregrounds different features of the same distribution, and moving between them allows the user to try to discover insights that a single view might have obscured.

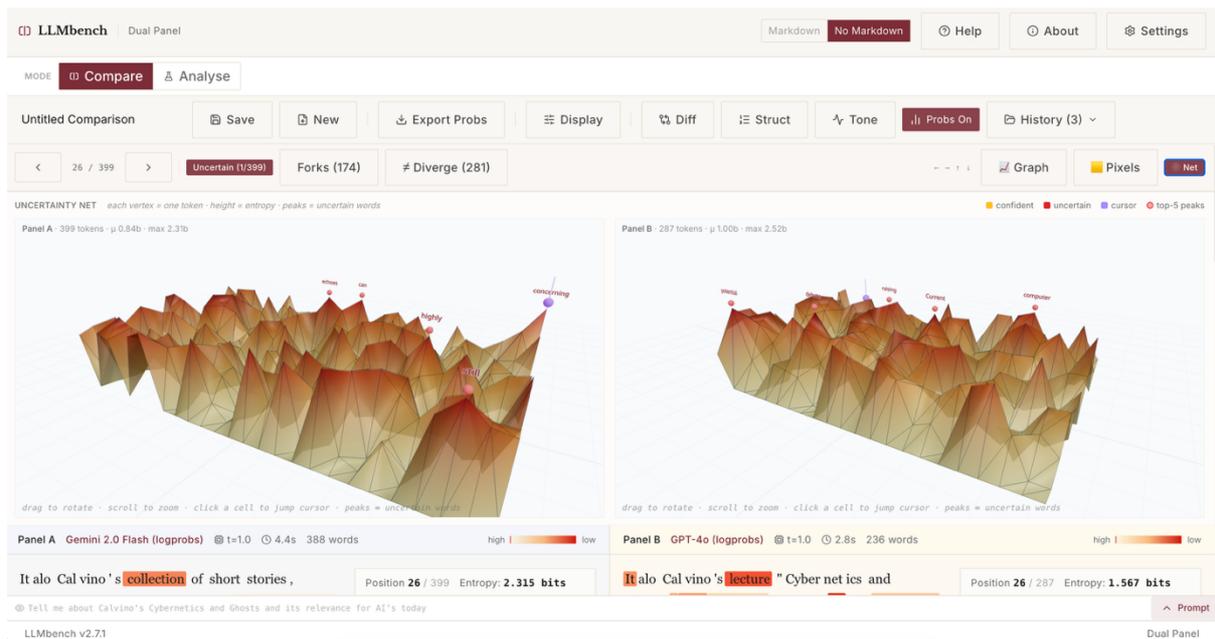

Figure 5: The Net band active, showing two 3D probability skyline meshes side by side, one for each panel. The WebGL terrain is visible with peaks corresponding to high-entropy token positions and flat areas corresponding to confident passages. Floating labels identify the top-5 highest-entropy points on each mesh. The standard text panels with heatmap overlay are visible below. The label bar at top identifies this as the Uncertainty Net.

## Diff, Struct, and Tone

The Probs family of views allows one to examine the model's internal probability distributions. The other three overlays in Compare mode work on the finished text as text, applying different analytical tools to make structural and rhetorical features visible.[4]

---

[4] There is an annotation system in Compare mode that supports six typed categories drawn from the Critical Code Studies methodology developed across Marino (2020) and the ELIZA reading project (Berry and Marino

The "Diff" overlay computes word-level differences between the two responses. Words present in one panel but absent from the other are highlighted, with unique-word counts appearing in each panel header. Both panels scroll in synchronisation so corresponding passages stay aligned. That the models said something different is obvious without any overlay, but the ability to diff across the same model (by selecting them in the settings panel) means that you can see how the same model generates different versions of the text. What the diff surfaces is the shape of that difference. Some models diverge lexically at the periphery, synonyms and minor phrasing variations. Others choose entirely different vocabularies for the same conceptual territory. The diff reveals which kind of divergence you are looking at.

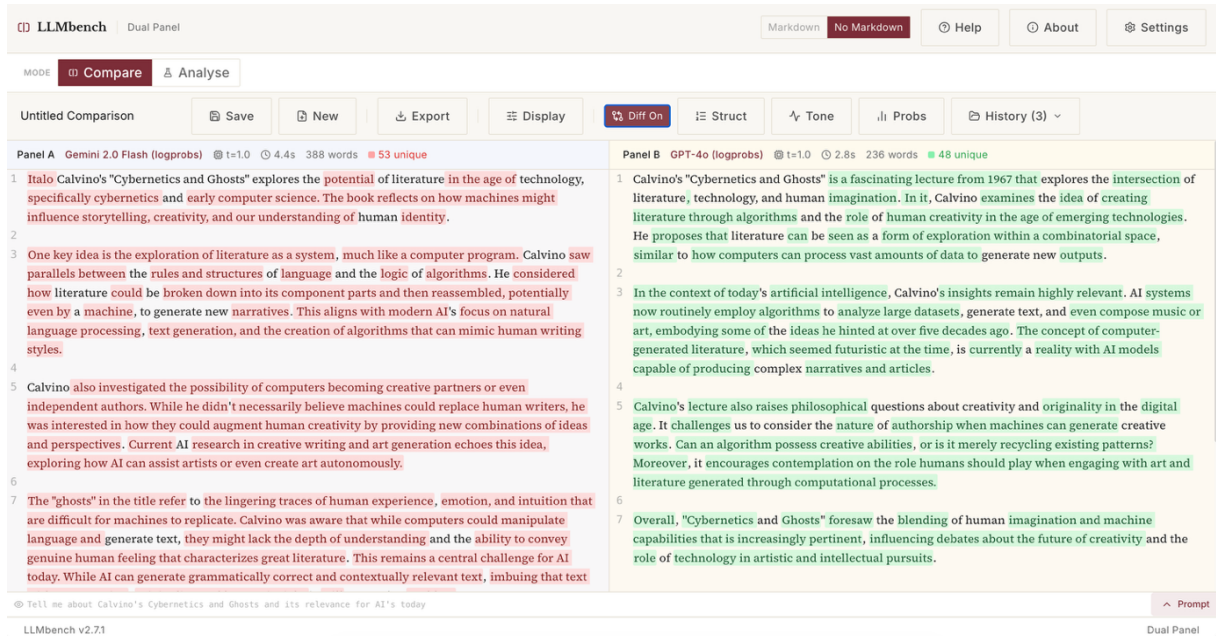

Figure 6: Compare mode with the Diff overlay active. Both panels display the Calvino responses with word-level highlighting. Words unique to each panel are highlighted in the respective panel's colour. Unique-word counts are visible in each panel header (Panel A: 52 unique; Panel B: 49 unique). The numbered sentence markers from the Struct view are visible in the gutter alongside the diff highlighting. Both panels are in synchronised scroll.

The "Struct" overlay presents the response as a sequence of numbered sentences, breaking the running prose into discrete structural units that can be inspected individually. Each sentence is tagged with a sentence number in the gutter and displayed on its own line, making the argumentative arc of the response visible. Discourse connectives (e.g. "however", "therefore", "furthermore", "consequently", "nevertheless", "moreover", and roughly twenty others) are highlighted with hover tooltips indicating the rhetorical function each serves at that position. What the Struct view surfaces is the structural grammar of the response. How long are the sentences on average? Where does the model pile up subordinate clauses, where does it open with a fragment, where does it use a discourse marker to pivot its argument? The overlay

---

2024), Observation, Question, Metaphor, Pattern, Context, and Critique. Select any text in either panel to activate the annotation widget. Annotations persist to browser localStorage and export with the comparison as JSON, plain text, or PDF with coloured annotation badges.

does not analyse these features automatically. It presents the text in a form that makes them visible to the reader.

The "Tone" overlay applies Ken Hyland's (2005) metadiscourse model to both generated texts. Seven register categories are applied across the text. These are, (1) Hedges, words like "might", "perhaps", "arguably", are in blue, (2) Boosters, "clearly", "certainly", "must", are in green, (3) Limiting terms, "not", "never", "without", are in orange, (4) Attitude markers, "important", "surprising", "problematic", are in purple, (5) Intensifiers, "very", "extremely", "highly", are in amber, (6) Self-mentions, "I", "we", "our", are in rose, and (7) Engagement markers, "you", "consider", "note", "imagine", are in teal.[5] The user can click any "chip" in the navigation bar to toggle a category. The question mark beside each chip opens the Hyland definition. Hover any marked word for its surrounding context, frequency count, and a brief linguistic note on its function at that position.

What the Tone view shows is the rhetorical stance each model adopts toward its own claims and toward its reader. A response dense with hedges is a cautious one. A response heavy with boosters is assertive, perhaps overconfident. When the Calvino prompt is sent to Gemini and GPT-4o, the distributions differ in ways that would be difficult to characterise without the overlay making them visible. The balance bar at the foot of each panel shows proportional category distribution, a summary of each model's rhetorical style.

Figure 7: Compare mode with the Tone overlay active. Both panels display the Calvino responses with Hyland's metadiscourse categories applied as colour-coded highlights throughout the text. The category count chips are visible at the top of each panel (Hedges, Boosters, Limiting, Attitude, Intensifiers, Self-mentions, Engagement), with numerical counts. The register balance bar at the foot of each panel shows proportional distribution across categories. Highlighted words in different colours are distributed throughout both responses.

---

[5] The Hyland (2005) metadiscourse framework was developed for the analysis of academic writing. Its application here to AI-generated text is an experimental approach in this tool. LLMs are not writing academic prose exactly, but they have been trained on vast quantities of it, and the patterns of hedging, boosting, and reader engagement that Hyland identifies are present in many model outputs and vary systematically across models, prompts, and training regimes.

## The Analyse Modes

Where Compare mode supports close reading of individual outputs, the five Analyse modes run empirical probes. Each poses a specific question about model behaviour and returns quantitative results. The most productive workflow would probably use both together, combining a close reading to identify what seems interesting about a given response and analytical modes to see how it is differently presented.

The "Stochastic Variation" mode sends the same prompt to the same model between three and twenty times in succession and reports how much the outputs differ from one another. This is among the more counterintuitive aspects of model behaviour, and observing it directly is very helpful for analysis. Temperature above zero means the model samples from its probability distribution rather than always selecting the highest-probability token, and sampling is by definition stochastic.

In the Calvino example, five runs of Gemini 2.0 Flash produce an average word count of 386, average vocabulary diversity of 54.8%, and average pairwise word overlap of 42.4%. Each run appears as a result card with its own metrics. The Deep Dive expands to show a pairwise overlap matrix across all runs, colour-coded from green for high overlap through yellow to red, giving a quick visual summary of where the model is consistent and where it is different.

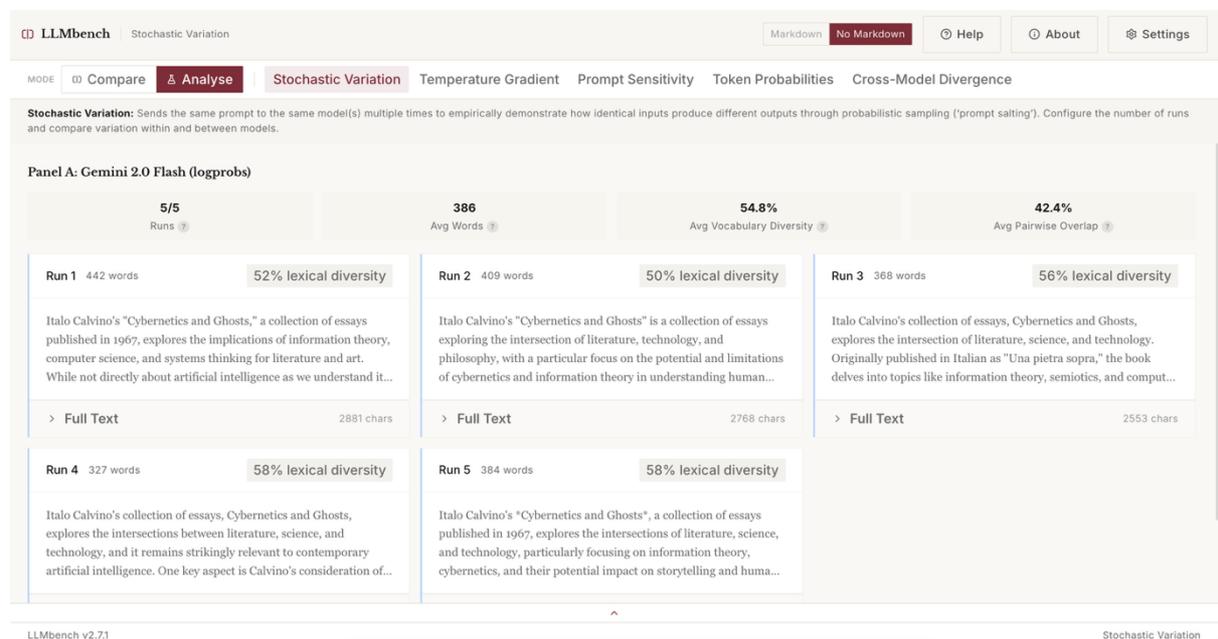

Figure 8: Stochastic Variation mode showing five runs of the Calvino prompt to Gemini 2.0 Flash. Summary statistics at the top: 386 average words, 54.8% average vocabulary diversity, 42.4% average pairwise overlap, 5/5 runs complete. Three result cards are visible in the main area: Run 1 (442 words, 52% lexical diversity), Run 2 (409 words, 50% lexical diversity), Run 3 (368 words, 56% lexical diversity), with Run 4 and Run 5 cards partially visible below. The word count and lexical diversity metrics are labelled under each card.

The "Temperature Gradient" mode runs the same prompt across six fixed sampling temperatures, from 0.0 through 0.3, 0.7, 1.0, 1.5, to 2.0. Temperature 0.0 is deterministic, always selecting the highest-probability token. At 2.0, sampling is randomised enough that the output begins to explore quite unlikely regions of the model's vocabulary. Reading the six result cards sequentially is a record of how randomness shapes output.

The Calvino prompt at temperature 0.0 produces 456 words with 48% lexical diversity; at 2.0, 382 words with 61% diversity. The Deep Dive provides a per-temperature metrics table covering word count, sentence count, average sentence length, vocabulary diversity, and response time, with contextual notes on how low- and high-temperature behaviour typically differs.

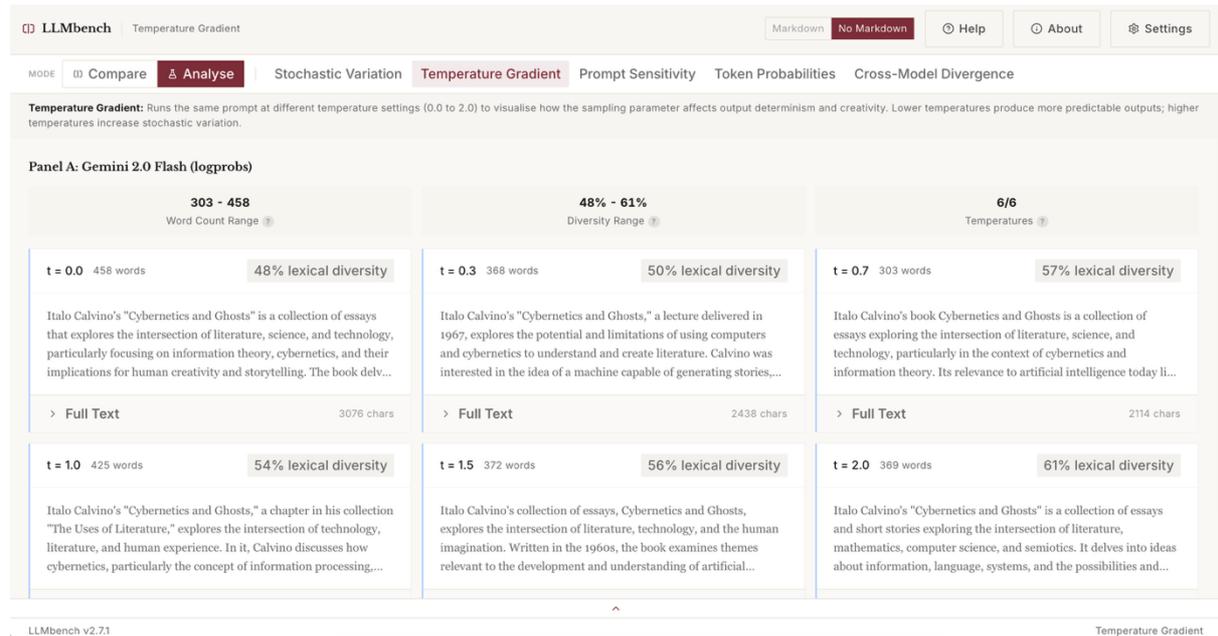

Figure 9: Temperature Gradient mode showing six results for Gemini 2.0 Flash at temperatures 0.0, 0.3, 0.7, 1.0, 1.5, and 2.0. The summary strip at the top shows word count range 303-458, diversity range 48-61%, and 6/6 temperatures complete. Six result cards are arranged in two rows of three. Each card shows temperature value, word count, lexical diversity percentage, and a preview of the response text. The Full Text expand link is visible beneath each preview.

"Prompt Sensitivity" tests how minor changes to a prompt affect model outputs. The mode auto-generates variations from the base prompt, adding "please", changing punctuation, rephrasing as a question, adding "step by step", and so on, then ranks each variation by its word overlap with the base output. Custom user-defined variations can be added to the set. The results reveal which prompt tweaks produce the largest divergence from the base response.

In the Calvino example, the prompt variations range from 33% overlap with the base (adding a period) to 41% overlap (adding step by step), with the question-form variant and the please variant sitting in between. This is a fast empirical check on prompt brittleness. If small phrasings produce large divergences, the responses being read in Compare mode should be treated as draws from a probability distribution.

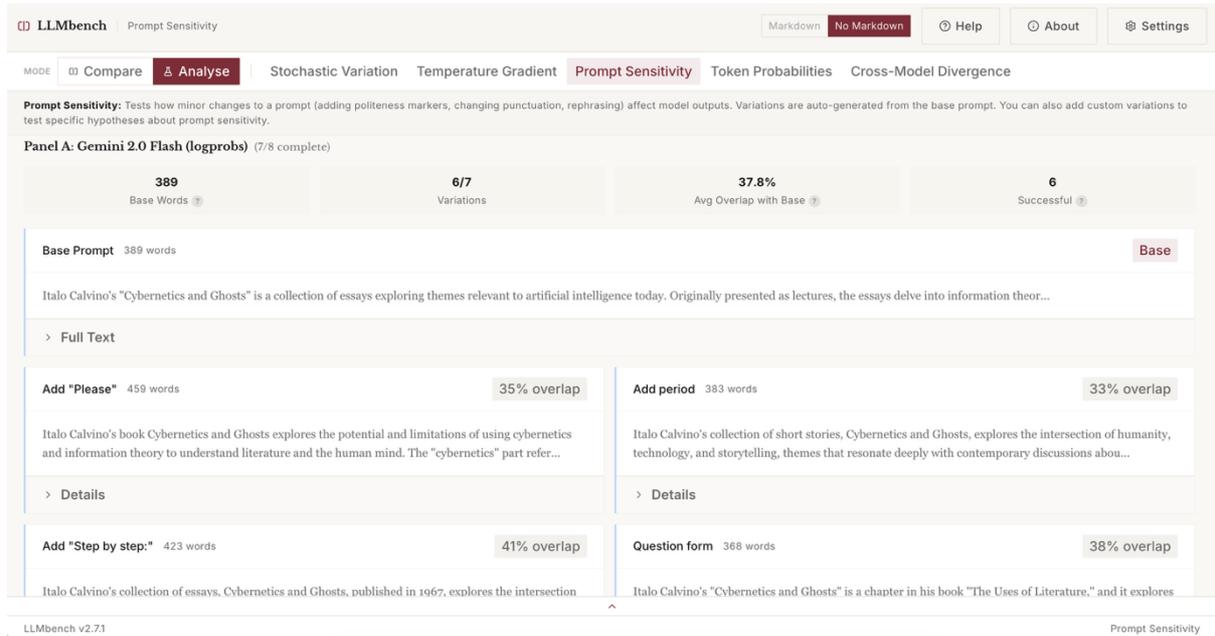

Figure 10: Prompt Sensitivity mode showing the base Calvino prompt and four auto-generated variations. The summary strip shows 389 base words, 6/7 variations complete, 37.8% average overlap with base, 6 successful runs. The Base Prompt card is visible at the top (389 words, labelled Base). Below it, four variation cards: Add "Please" (458 words, 35% overlap), Add period (383 words, 33% overlap), Add "Step by step" (423 words, 41% overlap), and Question form (369 words, 38% overlap). The Details expand link is visible under each card.

The "Token Probabilities" mode provides an environment for single-response "logprob" analysis. Where the Probs overlay in Compare mode is designed for comparative work across two responses, this mode is for extended inspection of a single output. The summary bar at the top reports mean entropy, average probability, the token with the maximum entropy, and total token count. Below it, an entropy distribution histogram divides all tokens into five confidence bands from "Very Low" to "Very High" and clicking any bar lists the exact tokens that fall into it. A Sentence Entropy view colour-codes each sentence by its mean token entropy, surfacing which sentences carry the most uncertainty at a structural level. The Uncertainty Deep Dive at the bottom provides hotspot lists and the most frequently considered alternatives across all positions.

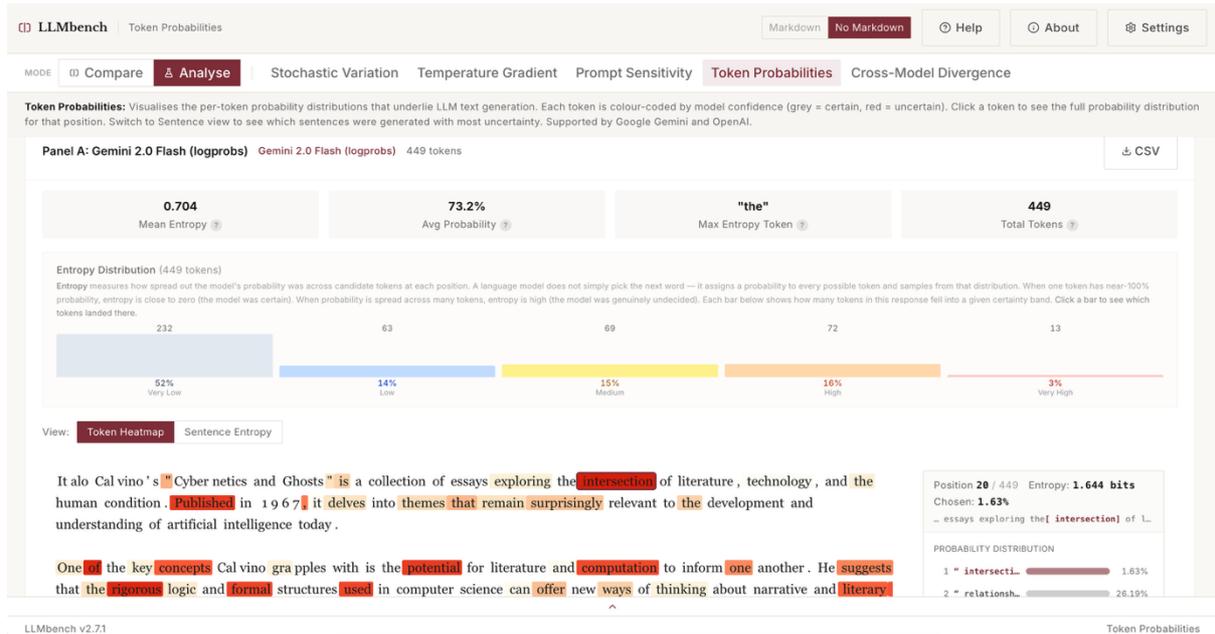

Figure 11: Token Probabilities standalone mode for Gemini 2.0 Flash responding to the Calvino prompt. The summary bar shows Mean Entropy 0.704, Avg Probability 73.2%, Max Entropy Token "the", Total Tokens 449. Below, the Entropy Distribution histogram is visible with five confidence bands. The Token Heatmap tab is active, showing the coloured response text with a probability inspector pinned to the right showing Position 26/449, Entropy 1.644 bits, Chosen 1.63% (intersects...), with probability distribution bars for the top alternatives. The Sentence Entropy tab selector is visible for switching views.

"Cross-Model Divergence" provides the quantitative frame for what Compare mode examines qualitatively. The headline metrics are Jaccard similarity, the proportion of unique words shared relative to the union of both vocabularies, and word overlap percentage. Structural metrics for each panel appear beneath, such as word count, sentence count, average sentence length, vocabulary diversity (note these results are from a different run from the Compare section).

In the Calvino comparison, Gemini 2.0 Flash produces 322 words across 16 sentences with average sentence length 20.1 and 58% vocabulary diversity. GPT-4o produces 262 words across 10 sentences, average sentence length 26.2, 63% vocabulary diversity. The Jaccard similarity between them is 22.8%, meaning the two models are drawing on largely different lexical territory despite responding to the same prompt. The Vocabulary Analysis Deep Dive partitions the vocabulary into what is unique to A, shared, and unique to B, alongside top-word frequency bar charts and unique bigram candidates for each panel.

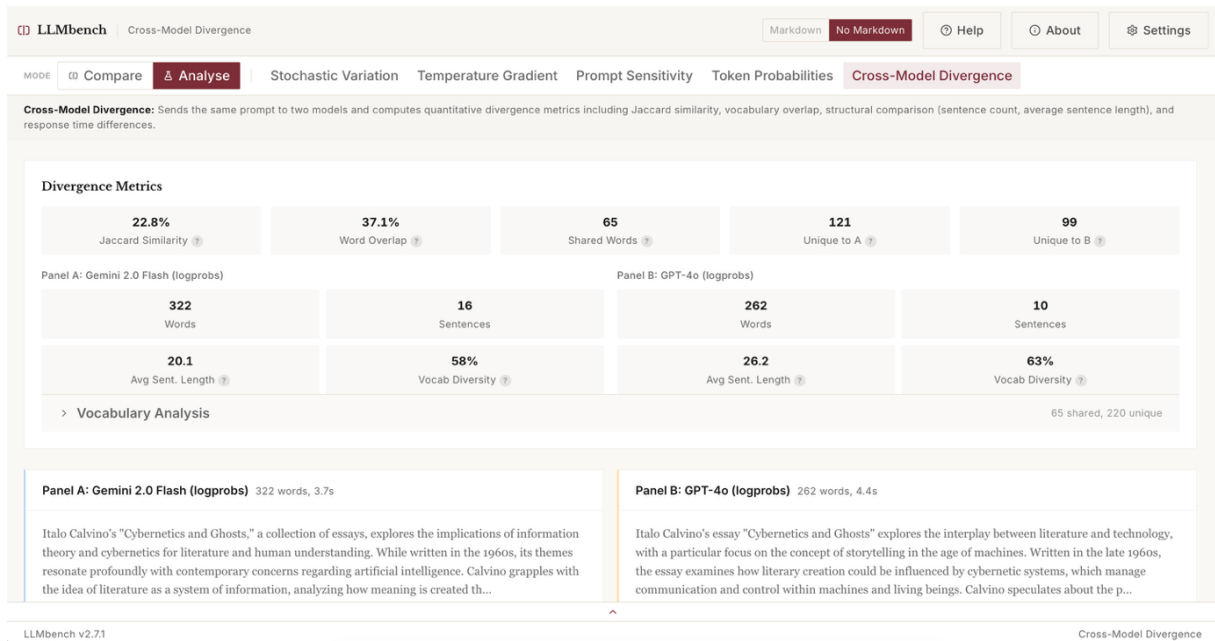

Figure 12: Cross-Model Divergence mode comparing Gemini 2.0 Flash and GPT-4o on the Calvino prompt. The Divergence Metrics panel shows: Jaccard Similarity 22.8%, Word Overlap 37.1%, Shared Words 65, Unique to A 121, Unique to B 99. Below, per-panel structural metrics: Panel A (Gemini 2.0 Flash) 322 words, 16 sentences, 20.1 avg sent length, 58% vocab diversity; Panel B (GPT-4o) 262 words, 10 sentences, 26.2 avg sent length, 63% vocab diversity. The Vocabulary Analysis expandable section is visible. Both full response text panels are visible at the bottom (note these results are from a different run from the Compare section).

## Example: Position 26 in the Calvino Comparison

Some of the more abstract claims above are perhaps most concretely demonstrated by working through a single position in the Calvino comparison. Figure 1 showed the two responses at a glance, both opening on Calvino's *Cybernetics and Ghosts* and both moving towards its relevance for AI. Position 26 in each token stream, shown in Figure 2, shows the two models in very different moments.

Gemini 2.0 Flash at position 26 is mid-sentence, extending a characterisation of the essay: "... explores the intersection of humanity and technology, particularly [concerning] themes that resonate strongly with contemporary discussions surrounding artificial intelligence." The token it has just chosen is "concerning", at 11.76% probability. That figure is lower than the probabilities of three alternatives the model also considered. Ranked by probability, the top five candidates are " the" (15.11%), " through" (13.91%), " in" (12.80%), " concerning" (11.76%), and " themes" (11.76%). Roughly 34.7% of the probability mass is distributed further down the tail. The model had no strong preference at this position, and temperature sampling produced "concerning" rather than the marginally more probable "the". Entropy is 2.315 bits, in the high range for this response.

What is interesting here is what the alternatives would have done. " through" would have produced "particularly through themes...", a slightly more conventional syntactic move with roughly the same semantic outcome. " in" would have shifted the register towards "particularly in themes..." and required a different continuation. " the", the most probable candidate, would have opened an entirely different grammatical structure, since "particularly the..." suggests the sentence would have moved to specify which themes, rather than to characterise them. The model's actual choice, "concerning", commits to a prepositional construction that the sentence will need to complete by naming what the themes concern, and indeed it does, "contemporary discussions surrounding artificial intelligence". A fork here would have changed more than the phrasing. The shape of the argument would have been different too.

GPT-4o at position 26 is in a different part of its response. Having opened by situating the essay historically ("... was delivered in 1967 at the University of Turin"), it is about to connect the Turin lecture to Calvino's later book *The Uses of Literature*. The chosen token is " and" at 49.27%. The alternatives are tighter, "." (37.19%), " as" (8.17%), "," (4.73%), " during" (0.36%). Entropy is 1.567 bits, substantially lower than Gemini's at the same position.

The two-token alternation here, " and" versus ".", is analytically revealing. " and" continues the sentence, staying with the biographical-historical register and connecting the Turin lecture to the later book. "." ends the sentence, after which a new sentence would need to restart the discursive frame. The model is roughly twice as likely to continue as to stop, but the stopping option is live. Everything below those two is a tail of much less probable continuations. This is a decisive moment, keep going or begin a new sentence, a choice closer to a binary than the Gemini position, which faced a spread of five near-equivalents.

That "position 26" lands the two models at different semantic-syntactic moments is itself a finding. The same ordinal position in two token streams is not the same discursive position, because the two models have structured their opening paragraphs differently. Gemini has moved quickly into characterising the essay's concerns, having spent its first 26 tokens describing the collection and stating a thesis about its "intersection". GPT-4o has stayed with biographical-historical framing, and by token 26 it is still locating the lecture in 1967 at Turin. The "Diverge" navigation chip surfaces 281 such positions across the two responses. Each is, on inspection, a different kind of discursive moment in each model, and the reading moves between surface comparison (what word is here versus what word is there) and distributional comparison (what shape of uncertainty is here versus there).

The numbers in Figure 2 (2.315 bits versus 1.567 bits, 11.76% versus 49.27%) are the material the analysis works with rather than the analysis itself. The interpretive work lies in asking what the distribution is doing at this moment, how close the runner-up was, whether the alternatives would have produced substantively different continuations, and how this moment of uncertainty relates to the model's broader rhetorical trajectory. The inspector makes those questions askable and explorable in useful ways.

## Limitations

Several limitations of the current system deserve acknowledgement, some technical, some methodological. The most significant technical constraint concerns log-probability access.

Not all model providers expose token-level probability data through their APIs, and those that do are not uniform in what they return. Google Gemini (2.0 and later) and OpenAI models return usable logprob data. Anthropic's Claude models, at the time of writing, do not expose this data through the public API. OpenRouter's routing layer surfaces logprobs for OpenAI models but not for most of the open-weight models it serves. This means the Probs overlay, which I have described as the analytical heart of the tool, is available only for a subset of the models that can be compared. Readers should treat the probabilistic analysis as applying to the accessible models rather than as a general capability across the LLM ecosystem. Different models use different tokenisation schemes, and a "token" in GPT-4o is not the same unit as a "token" in Gemini 2.0 Flash. When the tool reports that both models chose different tokens at "position 26", it is reporting a position in each model's own token sequence, not a position in a shared segmentation of the prompt and response. Cross-model divergence metrics at the token level therefore mix together genuine hermeneutic difference (the models really did choose different words) and tokenisation artefact (what counts as a token varies). The surface comparison (words, sentences) is more stable across models. The token-level comparison should be read with this caveat in mind.

Commercial APIs are moving targets. A model version today may behave slightly differently tomorrow, even if the nominal version string is unchanged, because providers routinely update weights, safety filters, and routing heuristics without publishing full changelogs. The comparisons generated by LLMbench are therefore historical artefacts, snapshots of model behaviour at the moment of the API call. Saved comparisons export prompt, outputs, logprob data, and provenance metadata (model, temperature, timestamp), which allows a reading to be re-examined later even if the model has changed. It does not allow the reading to be re-run as if nothing had changed in the intervening period.

A methodological limitation concerns the Hyland metadiscourse framework. As noted earlier, Hyland's categories were developed for the analysis of academic writing, and their application to LLM outputs is experimental. The categories appear to carry across because LLMs have been trained on enormous quantities of academic prose and the patterns of hedging, boosting, and reader engagement that Hyland identifies are present in much of that training data. But LLM output is not academic prose, and some of the register moves visible in model responses (the characteristic "as an AI language model" self-mentions, the bullet-list scaffolding, the particular rhetorical style of RLHF-tuned assistants) are not well described by Hyland's scheme. An alternative framework developed specifically for LLM rhetoric would be a useful complement.

Finally, the readings the tool enables are interpretive readings, not measurements. High entropy at a given token position tells us the model was uncertain among several alternatives. It does not tell us what the uncertainty meant, whether it reflects a semantically genuine fork in the argument, a syntactic decision point, a rote hedge the model learned to produce at certain moments, or simply an arbitrary choice among near-synonyms.

The tool surfaces the distribution. The interpretation of what the distribution is doing at a particular point remains the reader's work. This is a feature rather than a bug for a workbench designed for hermeneutic use, but it means the tool does not provide, and is not intended to provide, an automated analysis of what the probabilistic structure means at any given position.

## Conclusion

What LLMbench makes possible is a form of reading that is comparative but also not reducible to the numbers the engineering tools report.[6] The token probability views reveal the model's counterfactual history, every token is a road taken and the distribution shows the roads not taken. The high-entropy positions, the forks, the diverge points, are the moments where the text could plausibly have gone several different ways, where what appears in the panel as a settled word or phrase was, a fraction of a second before sampling, a genuinely open question. I argue that logprobs are an underutilised tool in humanistic and social scientific readings of AI, possibly the closest we get to seeing which parts of the text the model was committed to and which parts could just as easily have come out differently, where it was, we might say, effectively rolling dice within a set of similar words.

The variorum principle from *10 PRINT* (Montfort et al. 2013) is a key principle behind the ideas for this tool. Different variants of the same text, whether produced by different models or by the same model at different temperatures, are analytically productive to explore. They reveal what is deterministic and what is contingent, what depends on particular training decisions and what the models share. For example, the Analyse modes make the within-model variants visible, and the Compare mode makes the across-model comparison more visible and more amenable to close reading. Together they are a workbench for the comparative hermeneutics of AI-generated text, which, I would argue, can contribute to the scholarly practice of working with and critiquing generative AI models.[7] What such reading finally shows, to borrow from Ricoeur, is that the meaning of these texts is not hidden behind them, in benchmark scores or alignment metrics, but disclosed in front of them, in the cloud of near-equivalent alternatives the model passed over to arrive at the words it wrote.[8]

---

[6] A code-editor substrate (CodeMirror) is used to display prose text and discussed in the README. Briefly, the aim was to support a gutter and line-based annotation system taken from another tool I developed called CCS Workbench. The result is a reading environment positioned between a word processor and a code editor, which is a document analysis environment with a scholarly annotation system built in.

[7] The Vector Lab tools draw on a theoretical position I have been developing elsewhere (Berry 2025, Berry under review), that LLMs inhabit a high-dimensional vector space or manifold and that reading their outputs is partly a matter of reading against the structure of that vector space, not only against the surface of the text. The tools in the series are attempts to make aspects of that space legible to humanistic close reading.

[8] The tool is browser-based and requires only an API key from a supported provider. OpenRouter gives the broadest model access through a single key. The deployed version is at https://llm-bench-mu.vercel.app/. Code, documentation, and the full architecture and description are at https://github.com/dmberry/LLMbench. Contributions, issues, and forks are welcome. LLMbench is MIT licensed.

# Bibliography

Anthony, L. (2022) AntConc (Version 4.2.0) [Computer Software]. Tokyo: Waseda University. Available at: https://www.laurenceanthony.net/software.

Berry, D. M. (2025) 'Synthetic media and computational capitalism: towards a critical theory of artificial intelligence', *AI & Society*, 40(7), pp. 5257-5269. https://doi.org/10.1007/s00146-025-02265-2

Berry, D. M. (under review) 'Vector Theory'.

Berry, D. M. and Marino, M. C. (2024) 'Reading ELIZA: Critical Code Studies in Action', *Electronic Book Review*. https://electronicbookreview.com/essay/reading-eliza-critical-code-studies-in-action/

Chiang, W.-L., Zheng, L., Sheng, Y., Angelopoulos, A. N., Li, T., Li, D., Zhang, H., Zhu, B., Jordan, M., Gonzalez, J. E. and Stoica, I. (2024) 'Chatbot Arena: An Open Platform for Evaluating LLMs by Human Preference', arXiv preprint arXiv:2403.04132.

Gius, E., Meister, J. C., Meister, M., Petris, M., Bruck, C., Jacke, J., Schumacher, M., Gerstorfer, D., Flüh, M. and Horstmann, J. (2021) CATMA 6 [Computer Software]. Available at: https://catma.de.

Hyland, K. (2005) *Metadiscourse: Exploring Interaction in Writing*. Continuum.

Kahng, M., Tenney, I., Pushkarna, M., Liu, M. X., Wexler, J., Reif, E., Kallarackal, K., Chang, M., Terry, M., and Dixon, L. (2024) 'LLM Comparator: Visual Analytics for Side-by-Side Evaluation of Large Language Models', *IEEE Transactions on Visualization and Computer Graphics*. https://arxiv.org/abs/2402.10524

Liang, P. et al. (2023) 'Holistic Evaluation of Language Models'. arXiv. Available at: https://doi.org/10.48550/arXiv.2211.09110.

Marino, M. C. (2020) *Critical Code Studies*. MIT Press.

McCallum, A. K. (2002) MALLET: A Machine Learning for Language Toolkit. Available at: https://mallet.cs.umass.edu.

Montfort, N., Baudoin, P., Bell, J., Bogost, I., Douglass, J., Marino, M.C., Mateas, M., Reas, C., Sample, M. and Vawter, N. (2013) *10 PRINT CHR$(205.5+RND(1)); : GOTO 10*. MIT Press.

Pasquinelli, M. (2023) *The Eye of the Master: A Social History of Artificial Intelligence.* Verso.

Ricoeur, P. (1981) 'Metaphor and the central problem of hermeneutics', in *Hermeneutics and the Human Sciences*, ed. J. B. Thompson. Cambridge University Press, pp. 127-143.

Simon, R. et al. (2017) 'Linked Data Annotation Without the Pointy Brackets: Introducing Recogito 2', *Journal of Map & Geography Libraries*, 13(1), pp. 111–132. Available at: https://doi.org/10.1080/15420353.2017.1307303.

Sinclair, S. and Rockwell, G. (2016) *Voyant Tools.* https://voyant-tools.org/.

Srivastava, A. *et al.* (2023) 'Beyond the Imitation Game: Quantifying and extrapolating the capabilities of language models'. arXiv. Available at: https://doi.org/10.48550/arXiv.2206.04615.